\documentclass[showpacs,preprintnumbers,amsmath,amssymb]{revtex4}

\usepackage{graphicx}% Include figure files
\usepackage{dcolumn}% Align table columns on decimal point
\usepackage{bm}% bold math
\newcommand{\be}{\begin{equation}}
\newcommand{\ee}{\end{equation}}
\newcommand{\ben}{\begin{eqnarray}}
\newcommand{\een}{\end{eqnarray}}
\begin{document}

%\preprint{draft version}

\title{Metric character of the quantum Jensen-Shannon divergence}

\author{P.W. Lamberti$^{1}$, A.P. Majtey$^{2}$, A. Borras$^{2}$,
M. Casas$^{2}$ and A. Plastino$^{3}$} \affiliation{$^1$ Facultad
de Matem\'atica, Astronom\'{\i}a y F\'{\i}sica, Universidad
Nacional de C\'ordoba, Ciudad Universitaria, 5000 C\'ordoba, and
CONICET, Argentina
\\\\ $^2$Departament de F\'{\i}sica and IFISC,
Universitat de les Illes Balears, 07122 Palma de Mallorca, Spain \\\\
$^3$ Instituto de F\'{i}sica La Plata, Universidad Nacional de La
Plata and CONICET, Argentina}

\date{\today}
\begin{abstract}
In a recent paper, the generalization of the Jensen Shannon
divergence (JSD) in the context of quantum theory has been studied
(Phys. Rev. A {\bf 72}, 052310 (2005)). This distance between
quantum states has shown to verify several of the properties
required for a good distinguishability measure. Here we
investigate the metric character of this distance. More precisely
we show, formally for pure states and by means of a numerical
procedure for mixed states, that its square root verifies the
triangle inequality.
\end{abstract}

\pacs{03.67.-a; 03.67.Mn; 03.65.-w}

\maketitle

\section{\label{sec:intro}Introduction}
Fundamental physical theories are  formulated in terms of an
abstract space. This is the case of  Relativity Theory, Quantum
Mechanics (QM), Yang-Mills like theories, and every proposal for
Unified Field Theory. On each abstract space different structures
can be defined. For example topological, differentiable, affine
and metric structures are ubiquitous in  space-time models. A
prescription for measuring just how close two points of the
concomitant space are is what we mean here by a metric structure.
A more precise distinction between a distance and a metric will be
given below.

In principle each one of above mentioned structures can be defined
in an independent way. In Special Relativity theory the space-time
is the standard $\mathbb{R}^4$ manifold provided with the (fixed,
non-dynamical) Minkowskian metric. In General Relativity, instead,
the space-time is a differentiable four dimensional manifold where
the metric is given by the matter-energy distribution (throughout
the Einstein's field equations). In both cases the metric is
compatible with Lorentz's covariance. It is worth mentioning here
(as known since the pioneering works of Gauss and Riemann) that
the metric defines every geometrical property of a differentiable
manifold.

In QM the corresponding abstract space is a (finite or infinite
dimensional) Hilbert space $\cal{H}$. In its mathematical
formalism the states of a physical system $\cal{S}$ are
represented by operators (density operators) acting on $\cal{H}$.
More precisely the states of the system $\cal{S}$ are represented
by the elements of ${\cal{B}}({\cal{H}})_1^+$, that is, the set of
positive trace one operators on $\cal{H }$. The notion of a state
as a unit vector of $\cal{H}$ refers to the extremal elements of
${\cal{B}}({\cal{H}})_1^+$ ($\rho \;\varepsilon \;
{\cal{B}}({\cal{H}})_1^+$ is extremal if and only if it is
idempotent, $\rho^2 = \rho$). In this case $\rho$ is of the form
$|\varphi\rangle\langle\varphi|$ for some unit vector
$|\varphi\rangle \; \varepsilon \; \cal{H}$, and is called a pure
state.

In the case of a Hilbert space, the basic underlying structure is
that of a vectorial space provided with an internal product
$\langle \;|\; \rangle$ between elements of $\cal{H}$. From this
inner product several ways of measuring ``proximity" between  two
elements of $\cal{H}$ can be defined. For example, the Wootters's
distance
\begin{equation}
d_W(|\varphi\rangle,|\psi\rangle) \equiv
d_W(|\varphi\rangle\langle\varphi|,|\psi\rangle\langle\psi|) =
\arccos(|\langle\varphi|\psi\rangle|) \label{Wootters}
\end{equation}
is a very important one. On one side (\ref{Wootters}) represents
the angle between the (pure) states $|\varphi\rangle$ and
$|\psi\rangle$; on the other, it has to do with the statistical
fluctuations in the outcomes of measurements into the QM formalism
\cite{Wootters}. Finally,  (\ref{Wootters}) is invariant under
unitary evolution. Therefore, we can think of (\ref{Wootters}) as
a very natural distance between pure states in QM, in some sense
imposed by the quantum theory itself. A generalization of this
distance to mixed states have been studied by Braunstein and Caves
\cite{Braunstein}.

Before going on let us remind the reader of  a formal
distance-definition. Let $\mathbb{X}$ be an abstract set. A
function
$$d:\mathbb{X} \times \mathbb{X}\rightarrow \mathbb{R}$$ is a distance defined
over the set $\mathbb{X}$, if for every $x,y \;\epsilon
\mathbb{X}$, it satisfies the following properties:

\begin{eqnarray}
d(x,y) & > & 0\;\; for\;\; x\ne y\;\; and \;\; d(x,x) = 0 \label{distproper} \\
d(x,y) & = & d(y,x) \nonumber
\end{eqnarray}

\noindent If, for every $x,y,z \;\epsilon \mathbb{X}$, the
function $d$ also verifies the triangle inequality:
\begin{eqnarray}
d(x,y) + d(y,z) - d(x,z)\geq 0 \label{triangular}
\end{eqnarray}
it is said that $d$ is a \textbf{metric} for the space
$\mathbb{X}$. Incidentally, we mention that the function given by
(\ref{Wootters}) is a metric. However, only a few among all
distances between quantum states historically introduced in the
literature verify condition (\ref{triangular}).

The definition of distance between mixed quantum states is a topic
of permanent interest. This interest has been lately rekindled on
account of problems emerging in information theory
(QIT)\cite{Lindblad, Jozsa, Lee, Luo, Markham}. In introducing
distances between quantum states, different roads have been
traversed. We have already mentioned the case of the Wootters's
distance and its generalization, presented in \cite{Braunstein}.
Recently, a rather interesting approach has been advanced by Lee
et al. in reference \cite{Lee}. There these authors characterize
the degree of closeness of two states with regards to the
information that can be attained for each of them from a complete
set of mutually complementary measurements plus an invariance
criterium. The resulting distance-measure  is equivalent to the
Hilbert-Schmidt metric. Let us recall that this metric emerges
from the  primitive structure of the Hilbert space. Indeed, an
inner product between bounded operators acting over the Hilbert
space $\cal{H}$ can be defined in the fashion
\[
\langle A| B \rangle =Tr(A^{\dag} B)
\]
The Hilbert-Schmidt norm of the operator $A$ is given by
$\|A\|_{HS}^2 = \langle A|A\rangle$ and from this, the
Hilbert-Schmidt metric between two operators $A$ and $B$ is
defined as
\begin{equation}
d_{HS}(A,B) = \|A - B \|_{HS}
\end{equation}

Another way of dealing with the problem of introducing distances
between quantum states is generalizing the notions of  distance
defined in the space of classical probability distributions. This
is the case of the relative entropy, which is a generalization of
information theoretic Kullback-Leibler divergence. The relative
entropy of an operator $\rho$ with respect to an operator
$\sigma$, both belonging to ${\cal{B}}({\cal{H}})_1^+$, is
\begin{equation}
S(\rho,\sigma)= Tr[\rho(\log\rho-\log\sigma)],
\end{equation}
where $\log$ stands for logarithm in base two. The relative
entropy is not a distance (and obviously is not a metric either)
because it is not symmetric  and does not verify the triangle
inequality (\ref{triangular}). Worst, it may even be unbounded. In
particular, the relative entropy is well defined only when the
support of $\sigma$ is equal to or larger than that of $\rho$
\cite{Lindblad} (the support of an operator is the subspace
spanned by the eigenvectors of the operator with nonzero
eigenvalues). This is a strong restriction which is violated in
some physically relevant situations, as for example when $\sigma$
is a pure reference state.

To overcome such problems we have recently investigated an
alternative to the relative entropy \cite{Majtey} that  emerges as
a natural extension of a symmetrized version of the
Kullback-Leibler divergence to the realm of quantum theory. In the
classical context this quantity is known as the Jensen-Shannon
divergence (JSD) and  was introduced by C. Rao \cite{Rao} and,
independently, by J. Lin \cite{Lin}. It has been applied to  a
diversity of problems arising in statistics  and physics
\cite{Roldan, Lamberti1, Lamberti2, Rosso, Crooks}. Among its most
significant properties one can include its boundedness and its
metric character \cite{Endres}. In reference \cite{Lamberti1} it
is shown that the JSD can be taken as a unifying distance between
probability distributions.

In our previous study of the quantum JSD we showed that it
verifies all the properties required for a good measure of
distinguishability between quantum states. In this paper we
investigate the metric property of the quantum JSD (QJSD), that
could be regarded as essential to check on the convergence of
iterative algorithms in quantum computation \cite{Galindo}.

The structure of this paper is as follows: next Section is devoted
to the formal definition of the classical and QJSD. In Section III
we investigate the metric character of the QJSD. In the first
place we consider  the pure states case and then we investigate
the metric properties for arbitrary mixed states recourse to
numerical simulations in different Hilbert spaces. Finally,  some
conclusions are  drawn in Sect IV.

\section{Classical Jensen-Shannon divergence and its quantum extension}

The classical JSD between two (discrete) probability distributions
$P=(p_1,p_2,...,p_N)$ and $Q=(q_1,q_2,...q_N)$, $\sum_i p_i =
\sum_i q_i =1$ is defined as

\begin{equation} D_{JS}(P,Q)=\frac{1}{2} \left[
S\left(P,\frac{P+Q}{2}\right)+S\left(Q,\frac{P+Q}{2}\right)
\right] \label{clasJSD}
\end{equation}
where $S(P,Q)=\sum_i p_i \log \frac{p_i}{q_i}$ is the
Kullback-Leibler divergence. $D_{JS}(P,Q)$ can be also expressed
in the form
\begin{eqnarray}
D_{JS}(P,Q) &= &H \left(\frac{P+Q}{2}\right) - \frac{1}{2} H(P) -
\frac{1}{2} H(Q) \nonumber \\
&=& \frac{1}{2} \left[\sum_i p_i \log \left(\frac{2p_i}{p_i+q_i}
\right) + \sum_i q_i \log \left( \frac{2q_i}{p_i + q_i} \right)
\right] \label{explicit1}
\end{eqnarray}
where $H(P) = - \sum_i p_i \log p_i$ is the Shannon entropy. The
classical JSD exhibits several interesting properties. Among them
we recall the following ones
\begin{itemize}
\item $D_{JS}(P,Q)$ is symmetric and always well defined;
\item it is bounded
\[
0\leq D_{JS}(P,Q)\leq1,
\]
\end{itemize}
and, as it was already stated,
\begin{itemize}
\item its square root,
\begin{equation}
d_{JS}(P,Q) \equiv \sqrt{D_{JS}(P,Q)} \label{sqrjs}
\end{equation}
verifies the triangle inequality Eq. (\ref{triangular}) (but
$D_{JS}$ does not).
\end{itemize}
A proof of this last fact can be found in references \cite{Endres,
Oster}. Alternatively, this can be proved by using some results of
 harmonic analysis due to I. Schoenberg \cite{Schoe}. The basic
fact that makes Schoenberg's theorem applicable to the classical
JSD resides in that it is a definite negative kernel, that is, for
all finite collection of real numbers $(c_i)_{i \leq N}$, and for
all corresponding probability distributions ${(P_i)}_{i \leq N}$,
the implication
\begin{equation}
\sum_{i=1}^N c_i=0 \Rightarrow \sum_{i,j} c_i c_j D_{JS}(P_i,P_j)
\leq 0 \label{schoe-cond}
\end{equation}
is valid. A corollary of  Schoenberg's theorem allows one to
assert that the probability distributions-space, with the metric
(\ref{sqrjs}), can be isometrically mapped into a subset of a
Hilbert space \cite{Topsoe}.

The classical JSD can be used to distinguish two probability
distributions and therefore can be used as well to do so for  two
quantum states described by their density operators, say, $\rho$
and $\sigma$. Indeed, let us suppose we choose a positive operator
value measure (POVM), $\sum_{i=1}^M \mathbb{E}_i = \mathbb{I}$,
that generates two probability distributions via
\begin{eqnarray}
p_i & = & Tr(\mathbb{E}_i \rho) \nonumber \\
q_i & = & Tr(\mathbb{E}_i \sigma), \nonumber
\end{eqnarray}
for $i=1,...,M$. Then we can use the JSD (\ref{clasJSD}) to
distinguish between these two distributions. In this procedure we
have the freedom of choosing the POVM which most clearly
distinguishes $p_i$ from $q_i$, that is, which makes the value of
$D_{JS}(p_i,q_i)$ the largest. This reasoning motivates to
introduce the quantity
\begin{equation}
D_{JS1}(\rho,\sigma)=\sup_{\{\mathbb{E}_i \}} D_{JS}(p_i,q_i),
\end{equation}
where the supremum is taken over all POVM's. Physically $D_{JS1}$
gives the best discrimination between the states $\rho$ and
$\sigma$ that we can achieve by means of measurements.

By mimicking the extension of Kullback-Leibler divergence to the
realm of quantum theory, we define the QJSD as \cite{Majtey}
\begin{equation}
D_{JS}(\rho,\sigma) =
\frac{1}{2}\left[S\left(\rho,\frac{\rho+\sigma}{2}\right)+
S\left(\sigma,\frac{\rho+\sigma}{2}\right)\right],
\label{JS-relative-entropy}
\end{equation}
that can be recast in terms of the von Neumann entropy $H_N(\rho)
= - Tr(\rho \log \rho)$ in the fashion
\begin{equation}
D_{JS}(\rho,\sigma) = H_N\left(\frac{\rho +
\sigma}{2}\right)-\frac{1}{2} H_N(\rho) - \frac{1}{2} H_N(\sigma),
\label{JS-vN}
\end{equation}
This quantity is always well defined, symmetric, positive definite
and bounded ($0\leq D_{JS}(\rho,\sigma) \leq 1$). By using the
corresponding properties of the relative entropy \cite{donald} and
expression (\ref{JS-relative-entropy}) it can be shown that, for
arbitrary $\rho$ and $\sigma$, the following inequality
\begin{equation}
D_{JS}(\rho,\sigma) \geq D_{JS1}(\rho,\sigma), \label{upper}
\end{equation}
is valid. The equality is satisfied \textit{if and only if} $\rho$
and $\sigma$ commute, that is, the upper bound in (\ref{upper})
is, in general, not attainable for any POVM.

To conclude this section we give the explicit expression for the
QJSD in terms of the eigenvalues and eigenvectors of the operators
involved in its expression.
\begin{equation}
D_{JS}(\rho,\sigma)= \frac{1}{2} \left[ \sum_{k,i} |\langle
t_k|r_i\rangle|^2 r_i \log \left(\frac{2r_i}{\tau_k} \right) +
\sum_{k,j} |\langle t_k|s_j \rangle|^2 s_j \log
\left(\frac{2s_j}{\tau_k} \right) \right], \label{explicit}
\end{equation}
where $\rho=\sum_i r_i |r_i \rangle \langle r_i|$, $\sigma=\sum_i
s_i |s_i\rangle\langle s_i|$, $(\rho+\sigma) = \sum t_i
|t_i\rangle\langle t_i |$ and $\tau_k =\sum_i r_i |\langle
t_k|r_i\rangle|^2 + \sum_i s_i |\langle t_k |s_i\rangle|^2$

It should be noted that, when $\rho$ and $\sigma$ do not commute,
the structure of (\ref{explicit}) is quite different from that of
(\ref{explicit1}).

\section{The metric character of the quantum $\sqrt{D_{JS}}$}

In this section we investigate the putative metric character of
the QJSD, that is we try to ascertain whether the square root of
the QJSD,
\begin{equation}
d_{JS}(\rho,\sigma)=\sqrt{D_{JS}(\rho,\sigma)}\label{dJS}
\end{equation}
verifies the triangle inequality. The other three properties for a
metric are obviously verified by (\ref{dJS}). A formal proof of
property (\ref{triangular}) for $\sqrt{D_{JS}(\rho,\sigma)}$ has
until now eluded us.  Unfortunately there is no analog of
Schoenberg's theorem when operators are involved. Still more,
there is no direct way of verifying condition (\ref{schoe-cond})
for expression (\ref{explicit}). No extension to the case of the
QJSD of the proof given in \cite{Endres} has been possible.
Incidentally it should be observed that, if the upper bound in
(\ref{upper}) could be attained for some POVM, the proof of the
triangle inequality for (\ref{dJS}) would be obvious (because
$\sqrt{D_{JS1}}$ verifies it).

The results to be presented here correspond to a separate analysis
of the metric condition for (\ref{dJS}) for the two cases: when
(\ref{dJS}) is restricted to pure states and when it acts on the
complete set ${\cal{B}}({\cal{H}})_1^+$. In the first instance we
were able to give a formal proof of inequality (\ref{triangular});
in the second one, we checked it by means of a numerical
algorithm.

\subsubsection{Pure states}
For a pure state the von Neumann entropy vanishes. Then, for two
pure states,
\begin{equation}
\rho=|\psi\rangle\langle\psi|\;\;\;and\;\;\;
\sigma=|\varphi\rangle\langle\varphi|,
\end{equation}
the QJSD (\ref{JS-vN}), becomes
\begin{equation}
D_{JS}(\rho,\sigma)=H_N\left(\frac{\rho + \sigma}{2}\right)
\label{pure}
\end{equation}
After some algebra, we can rewrite (\ref{pure}) in terms of the
inner product $\langle \; | \; \rangle$:
\begin{eqnarray}
D_{JS}(\rho,\sigma) &=& \Phi(|\langle \psi | \varphi \rangle|)
\equiv
\nonumber \\
 &=&-\left(
\frac{1-|\langle\psi|\varphi\rangle|}{2}\right)
\log\left(\frac{1-|\langle\psi|\varphi\rangle|}{2}\right) -\left(
\frac{1+|\langle\psi|\varphi\rangle|}{2}\right)
\log\left(\frac{1+|\langle\psi|\varphi\rangle|}{2}\right).
\label{entro2}
\end{eqnarray}

The entropy of the average $\frac{1}{2} \left(|\psi\rangle \langle
\psi| + |\varphi\rangle \langle\varphi| \right)$ can be
interpreted to the light of quantum information theory. Indeed,
let us suppose that Alice has a source of pure qubit signal states
$|\psi\rangle$ and $|\varphi\rangle$. Each emission is chosen to
be $|\psi\rangle$ or $|\varphi\rangle$ with an equal prior
probability  a half. Then the density matrix of the source is
$\frac{1}{2} \left(|\psi\rangle \langle \psi| + |\varphi\rangle
\langle\varphi| \right)$. Alice may  communicate the sequence of
states to Bob by transmitting one qubit per emitted state. But
according to the quantum source coding theorem, (\ref{pure}) gives
the lowest number of qubits per states that Alice needs to
communicate the quantum information (with arbitrarily high
fidelity) \cite{Schum}.

Let us take two fixed arbitrary pure states
$\rho=|\psi\rangle\langle \psi|$ and $\sigma =
|\varphi\rangle\langle \varphi|$ and an arbitrary third one, $\xi
=|\chi \rangle\langle \chi|$. Denote the absolute value of the
inner products $|\langle \psi|\varphi\rangle|$, $|\langle
\psi|\chi\rangle|$ and $|\langle \chi|\varphi\rangle|$ with $x,y$
and $z$, respectively, and  introduce then the function
\[
G(x,y,z) =\sqrt{\Phi(y)} + \sqrt{\Phi(z)} - \sqrt{\Phi(x)}.
\]
In terms of these variables the triangle inequality for
(\ref{dJS}) reads:
\begin{equation}
0\leq G(x,y,z). \label{hineq}
\end{equation}
We can decompose the vector $|\chi\rangle$ into i) a part
belonging to the plane determined by $|\psi\rangle$ and
$|\varphi\rangle$ and ii) another part perpendicular to that
plane:
\[
|\chi\rangle = a |\psi\rangle + b |\varphi\rangle +
|\chi_{\bot}\rangle.
\]
with $|a|\leq 1$ and $|b|\leq 1$. Then
\[
y = |a + b\langle \psi|\varphi\rangle| \;\; and \;\; z
=|a^*\langle \varphi|\psi\rangle + b^*|.
\]
As a function of $a$ and $b$, for $x$ fixed, $G$ is a concave
function on the circles $|a|\leq 1$ and $|b|\leq 1$ (in the sense
that its second derivative is negative) and it vanishes for $y=x$
and $z=x$. This guarantees that inequality (\ref{hineq}) is
satisfied for arbitrary $y$ and $z$.

\subsubsection{Arbitrary mixed states}

Here we attempt a numerical verification of the triangle
inequality for the distance (\ref{dJS}) when arbitrary mixed
states are involved. As a first approach, we numerically evaluate
the inequality (\ref{triangular}) by generating random states in a
N-dimensional Hilbert space. The space of all (pure and mixed)
such states can be regarded as a product space of the form
\cite{BATPLA06}:
\[ {\cal{H}}=\mathcal{P} \times \Delta,
\]
where $\mathcal{P}$ stands for the family of all complete sets of
ortho-normal projectors $\{\hat P_i\}_i^N,\,\,\sum_i \hat
P_i=\mathbb{I}$ ($\mathbb{I}$ the identity matrix), and $\Delta$
is the set of all real $N-$tuples of the form
$\{\lambda_1,\ldots,\lambda_N\};\,\,\lambda_i
\in\mathbb{R};\,\,\sum_i \lambda_i=1;\,\,0\le \lambda_i \le 1.$
Any state in  $\cal{H}$ is of the form $\rho=\sum_i \lambda_i \hat
P_i.$

In exploring exhaustively ${\cal{H}}$ we need to introduce an
appropriate measure $\mu $ on this space. Such a measure is
required to compute volumes within ${\cal{H}}$, as well as to
determine what is to be understood by a uniform distribution of
states on ${\cal H}$.  The measure that we adopt here is taken
from the work of Zyczkowski {\it et al.} \cite{ZHS98,Z99}.

An arbitrary (pure or mixed) state $\rho$ of a quantum system
described by an $N$-dimensional Hilbert space can always be
expressed as a product of the form:

\be \label{udot} \rho \, = \, U D[\{\lambda_i\}] U^{\dagger}. \ee

\noindent Here $U$ is an $N\times N$ unitary matrix and
$D[\{\lambda_i\}]$ is an $N\times N$ diagonal matrix whose
diagonal elements are, precisely, our above defined  $\{\lambda_1,
\ldots,\lambda_N \}$. The group of unitary matrices $U(N)$ is
endowed with a unique, uniform measure, known as the Haar's
measure, $\nu$ \cite{PZK98}. On the other hand, the $N$-simplex
$\Delta$, consisting of all the real $N$-uples $\{\lambda_1,
\ldots, \lambda_N \}$ appearing in (\ref{udot}), is a subset of a
$(N-1)$-dimensional hyperplane of $\mathbb{R}^N$. Consequently,
the standard normalized Lebesgue measure ${\cal L}_{N-1}$ on
$\mathbb{R}^{N-1}$ provides a measure for $\Delta$. The
aforementioned measures on $U(N)$ and $\Delta$ lead to a measure
$\mu $ on the set ${\cal S}$ of all the states of our quantum
system \cite{ZHS98,Z99,PZK98}, namely,

\be \label{memu} \mu = \nu
\times {\cal L}_{N-1}. \ee

In our numerical computations we randomly generate mixed states
according to the measure (\ref{memu}). In order to assess, for
these randomly generated states, how the triangle inequality
(\ref{triangular}) is satisfied, we define the auxiliary quantity

\begin{equation}
\Delta d_{JS}(\rho,\xi,\sigma)=d_{JS}(\rho, \xi)+d_{JS}(\xi,
\sigma)-d_{JS}(\rho,\sigma)
\label{delta}
\end{equation}
and evaluate it for a large enough number of simulated states.
This procedure is repeated for different dimensions of the Hilbert
space.

We investigate the positivity of $\Delta d_{JS}$, upon which the
metric character of the square root of the QJSD is based, by
constructing the probability distributions $\cal{P}$  for the
values of $\Delta d_{JS}$. The corresponding histograms, for
different dimensions of the Hilbert space, are depicted in Fig. 1.
As we are mainly interested in the positivity of $\Delta d_{JS}$,
we just plot the tails of the concomitant distributions $\cal{P}$,
selecting the portion for which one has, say,  $\Delta d_{JS} \, <
\, 0.2$. Such a choice allows us to portray  in sufficient  detail
the region of the distribution where a violation of the inequality
(\ref{triangular}) can  be detected.

\begin{figure}
\begin{center}
\includegraphics[scale=1.0]{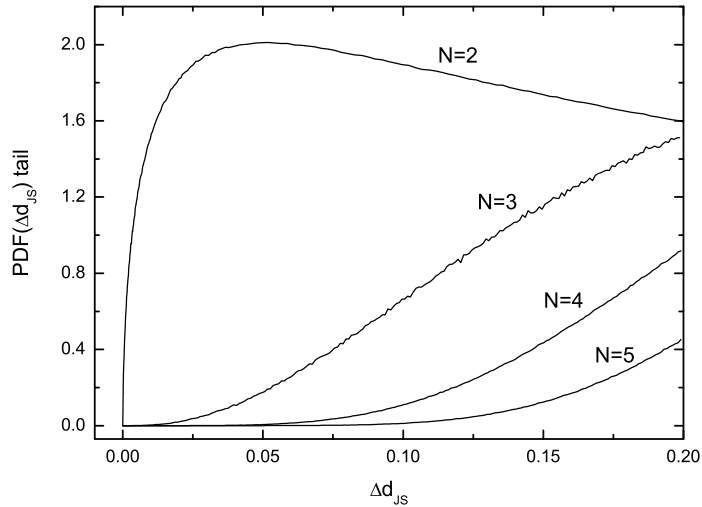}
\vskip -8mm \caption{Probability distribution for $\Delta d_{JS}$
for different Hilbert space dimensions. We just plot the tails of
the distributions for $\Delta d_{JS} \, < \, 0.2$. The tails were
constructed using of order of $10^7$ for $N=2$ and $10^6$ for
$N=3,4,5$ generated states.}
\end{center}
\label{Fig1}
\end{figure}

The probability for the particular value $\Delta d_{JS}=0$
actually represents the probability for finding a triplet of
density matrices for which $\Delta d_{JS} \leq 0$. None such
triplet of states has been found, which entails that the
probabilities for  violating the triangular inequality vanish for
all the distinct  Hilbert-space dimensions we have considered
here. Actually, the probability for low values of $\Delta d_{JS}$
becomes significantly smaller as the dimension of the pertinent
Hilbert space under study augments (the PDF's for higher Hilbert
space dimensions than those here reported have been also
computed).

The total number of randomly generated states was rather large ($
10^8$) in order to obtain a sufficiently large number of  points
belonging  to the tail-regions. These points fall then within the
zone of  low probabilities. The fact that no triplet of states
violating inequality (\ref{triangular}) has been encountered could
be thought of as being numerical evidence for the metric character
of the square root of the QJSD. The distributions in Fig. 1
clearly depend on the measure (\ref{memu}) used to compute them.
Higher probabilities for low values of $\Delta d_{JS}$ can
actually be obtained if one restricts the computation of the
histograms to states with a high degree of mixedness, although it
must be noted that such probabilities still diminish as the
dimension of the associated Hilbert space grows.

To avoid a statistical dependence on the measure (\ref{memu}) we
propose an alternative numerical approach by performing a
numerical minimization of $\Delta d_{JS}$. Any quantum mixed state
is completely determined by a finite number $n_p$ of parameters
which depends on the dimension of the Hilbert space. To determine
the minimum possible value of $\Delta d_{JS}$, one needs to find
the optimal values for such parameters. To such an end we use a
\emph{simulated annealing} algorithm in which the parameters are
iteratively modified until  convergence to the optimal values is
reached.

After running this algorithm for different Hilbert space
dimensions and for different triplets of initials states, one
detects always convergence to the same solution

\be \min \{ \Delta d_{JS}(\rho,\xi,\sigma) \}=0. \label{deltamin}
\ee

The optimal situation is reached when $\rho$ and $\xi$ are equal.
In our numerical search these states are always found to coincide
with the maximally mixed state for the Hilbert space-dimension
considered in each case. It is actually not enough to minimize Eq.
(\ref{delta}), because we wish it to be a minimum for any of the
three different ways to order the three states. If we minimize the
average of those three possible orderings, the minimum is also
$\Delta d_{JS}(\rho,\xi,\sigma) =0$, and it is obtained when the
three states become the maximally mixed state.

This last method, although does not provide us with a formal proof
of the metric character of the square root of the QJSD for mixed
states, does yield a clear and strong evidence about the validity
of the  conjecture advanced in the initial part   of this paper,
that constitutes the leitmotif of this work.

\section{Conclusions}

The main purpose of this work was to investigate the metrical
property of the QJSD. We were able to show  that the square root
of the QJSD verifies the triangle inequality, giving to this
distance the character of a metric. Although we have proved this
claim (for mixed states) only by giving numerical evidences, we
believe that the cases here  analyzed are sufficiently
representative so as to render credible  the claim that  metric
properties are verified in general for the QJSD.

A second item deserves to be pointed out, which emerges from the
following two facts:
\begin{itemize}

\item On the one hand, we have showed that, when restricted to pure states, the square
root of the entropy of the average $\frac{1}{2} \left(|\psi\rangle
\langle \psi| + |\varphi\rangle \langle\varphi| \right)$ is a true
metric.

\item  On the other hand, a classical result from Uhlmann
\cite{Jozsa} asserts that the fidelity of states $\rho$ and
$\sigma$

\[
F(\rho,\sigma)=Tr \sqrt{\rho^{\frac{1}{2}}\sigma
\rho^{\frac{1}{2}}}
\]
can be expressed in the form
\begin{equation}
F(\rho,\sigma)=
\max_{|\psi\rangle,|\varphi\rangle}|\langle\psi|\varphi\rangle|
\label{jozsa}
\end{equation}
where the maximization is over all purifications $|\psi\rangle$ of
$\rho$ and all purifications $|\varphi\rangle$ of $\sigma$
\cite{Nielsen}.
\end{itemize}

These two facts motivate us to introduce an alternative metric for
arbitrary mixed states. Given two arbitrary mixed states $\rho$
and $\sigma$ we can define
\begin{equation}
d_H(\rho,\sigma) = \min_{|\psi\rangle,|\varphi\rangle} \sqrt{H_N
\left( \frac{|\psi\rangle \langle \psi| + |\varphi\rangle
\langle\varphi|}{2} \right)} \label{new}
\end{equation}
where the minimum is taken over all purification $|\psi\rangle$ of
$\rho$ and all purifications $|\varphi\rangle$ of $\sigma$. In
(\ref{new}) we must look for  the minimum, not for the maximum as
in (\ref{jozsa}), due to the decreasing nature of $\Phi$, eq.
(\ref{entro2}), as a function of $|\langle\psi|\varphi\rangle|$.

Obviously the basic properties required for a good
distinguishability measure are inherited by (\ref{new}) from those
verified by the QJSD. Additionally, several interesting questions
arise from this proposal. For example, what relations exist
between  (\ref{new}) and (\ref{JS-relative-entropy}); or, in
general, how  (\ref{new}) relates to other quantum distances. A
more detailed study of the properties of this quantity will be
presented elsewhere.

\begin{acknowledgements}

This work was partially supported by the MEC grant FIS2005-02796
(Spain) and FEDER (EU) and by CONICET (Argentine Agency). AB and
APM acknowledge support from MEC through FPU grant AP-2004-2962
and contract SB-2006-0165. PWL wants to thank SECyT-UNC
(Argentina) and CONICET for financial support.
\end{acknowledgements}

\end{document}